\documentclass[conference]{IEEEtran}
\IEEEoverridecommandlockouts
\usepackage{cite}
\usepackage{balance}
\usepackage{amsmath,amssymb,amsfonts}
\usepackage{amsthm}
\usepackage{amssymb}
\usepackage{mathrsfs}
\usepackage{algorithm}
\usepackage{algpseudocode}
\usepackage{pifont}
\usepackage{adjustbox}
\usepackage{setspace}
\usepackage{graphicx}
\usepackage{textcomp}
\usepackage{xcolor}
\def\BibTeX{{\rm B\kern-.05em{\sc i\kern-.025em b}\kern-.08em
    T\kern-.1667em\lower.7ex\hbox{E}\kern-.125emX}}
\begin{document}

\title{BETA-UAV: Blockchain-based Efficient Authentication for Secure UAV Communication}
\vspace{-0.5em}
\author{\IEEEauthorblockN{Sana Hafeez,
Mahmoud A. Shawky,
Mohammad Al-Quraan, Lina Mohjazi, Muhammad Ali Imran and Yao Sun}\vspace{-0.3em}
\IEEEauthorblockA{James Watt School of Engineering, University of Glasgow, G12 8QQ, United Kingdom\\ Email: {\{s.hafeez.1}{\}@research.gla.ac.uk}.}
}
\maketitle

\begin{abstract}
Unmanned aerial vehicles (UAV), an emerging architecture that embodies flying ad-hoc networks, face critical privacy and security challenges, mainly when engaged in data-sensitive missions. Therefore, message authentication is a crucial security feature in drone communications. This paper presents a Blockchain-based Efficient, and Trusted Authentication scheme for UAV communication, BETA-UAV, which exploits the inherent properties of blockchain technology concerning memorability and is immutable to record communication sessions via transactions using a smart contract. The smart contract in BETA-UAV allows participants to publish and call transactions from the blockchain network. Furthermore, transaction addresses are proof of freshness and trustworthiness for subsequent transmissions.
Furthermore, we investigated their ability to resist active attacks, such as impersonation, replaying, and modification. In addition, we evaluate the gas costs associated with the functions of the smart contract by implementing a BETA-UAV on the Ethereum public blockchain. A comparison of the computation and communication overheads shows that the proposed approach can save significant costs over traditional techniques.
\end{abstract}
\section{Introduction}
Unmanned aerial vehicle (UAV) technology enhances the dependability and trustworthiness of transportation systems, particularly in heterogeneous and nonstationary data traffic scenarios. However, heterogeneous data sharing raises significant security and privacy concerns, preventing future intelligent transportation systems (ITS) from integrating UAVs\cite{b1}. Moreover, connectivity has become increasingly crucial in multiple-UAV systems. Drone communication is challenging because of (i) high node mobility, (ii) fluid topology, (iii) the long distance between nodes that can result in intermittent links, and (iv) power constraints.
Several features have contributed to the widespread use of UAV technology, such as coverage, exploratory possibilities, and intelligence-level rewards. Interest in UAVs is proliferating, and we can see that they are being deployed in many worldwide applications, such as aerial photography, agricultural production, and film and television production.
Establishing secure communication channels permits reliable UAV operations. However, external communication links, that is, between UAVs and infrastructure, and intra-vehicle communication must be protected. In addition, UAVs must ensure that only authorized entities have access to their resources and that all their internal modules are authenticated to achieve device security\cite{b2}. As UAVs operate without human intervention, device-to-device authentication is essential. Before a UAV can access a ground control station (GCS), all modules must be authenticated. 
Blockchain technology can be used to create a distributed system in which entities can enter and verify blocks, thereby ensuring system integrity. However, because users can request data for flying drones directly from UAVs instead of servers, drones continue to lose or leak data during transmission. This situation determines the complexity of a scheme. Consequently, the transmitted data may be subjected to extensive computation, which raises the possibility of privacy leakage. Furthermore, revealing privileged information can result in transmission security breaches. Therefore, a lightweight Blockchain-based Efficient Authentication BETA-UAV scheme was proposed for secure UAV communication. The objective is to enable mutual authentication and freshness identification such that the UAV can establish secure communication channels. Proof-of-freshness or authentication protocols allow UAVs to integrate into these systems quickly and securely.
In this study, we propose to accomplish the above-mentioned goals by conceiving new strategies by combining elliptic curve cryptography (ECC) and a trusted authentication scheme. We present a BETA-UAV blockchain-based efficient authentication for secure UAV communication that promises how the BETA-UAV can resist attacks. The objective is to enable mutual authentication and freshness identification so that the UAV network can establish secure communication channels. Proof-of-freshness or authentication protocols allow UAVs to integrate into these systems with minimal hassle and maximum security.
 \section{Related Works}\label{RW}
Recently, several studies have investigated the field of UAV system authentication. For example, this study provides an authentication framework for a UAV network using blockchain, 5G, and SHA-256. According to the authors, the proposed framework is secure against various IoD attacks and outperforms other schemes in terms of the communication overhead and computational costs.
However, the computational costs of this study are still high because of the use of SHA-256, which is inappropriate for UAVs. Li \emph{et al.} [3] proposed a lightweight communication mechanism that is supposed to be safer and faster than SM4 CTR.
Lei \emph{et al.} [4] illustrated a lightweight protocol for secure communication based on a physical unclonable function that employs a light mac function for encryption. Khalid \emph{et al.} [5] presented a light authentication scheme based on a non-cloneable physical component, particularly for vehicle networks. This framework utilizes a low-power and low-computing-intensive symmetric-encryption method.
Our concept is that if authentication can be excluded from the UAV, processing consumption will be reduced, thereby increasing flight duration and range.  The authors of [8] highlighted an initial architecture for UAV ID-based authentication. RFID tags provide unique identifiers for UAV within a scheme. Temporary UAV identification that provides both IDs is used to generate cryptography keys to protect privacy during the authorization procedure.
In short, some protocols proposed in recent years are vulnerable to attacks, such as inadequate security, encryption key predicting risk, privacy breaches, and server emulation. As a future development, secure communication between drones and GCS should be established. However, some of their solutions contain the issues discussed in [8]–[9].
The authors of \cite{b12} applied blockchain, 5G, and elliptic-curve technologies. ECC cryptography provides a framework for the authentication of drones.
The authors claim that the proposed framework is secure against numerous IoD attacks and outperforms comparable strategies in terms of communication and computational costs. However, this approach is computationally expensive owing to ECC, which is incompatible with UAVs. 
\section{The BETA-UAV Scheme}\label{BETA-UAV}
This section examines the layout of the BETA-UAV scheme depicted in Figure 1.
\begin{figure}[t!]
\centerline{\includegraphics[width=7cm]{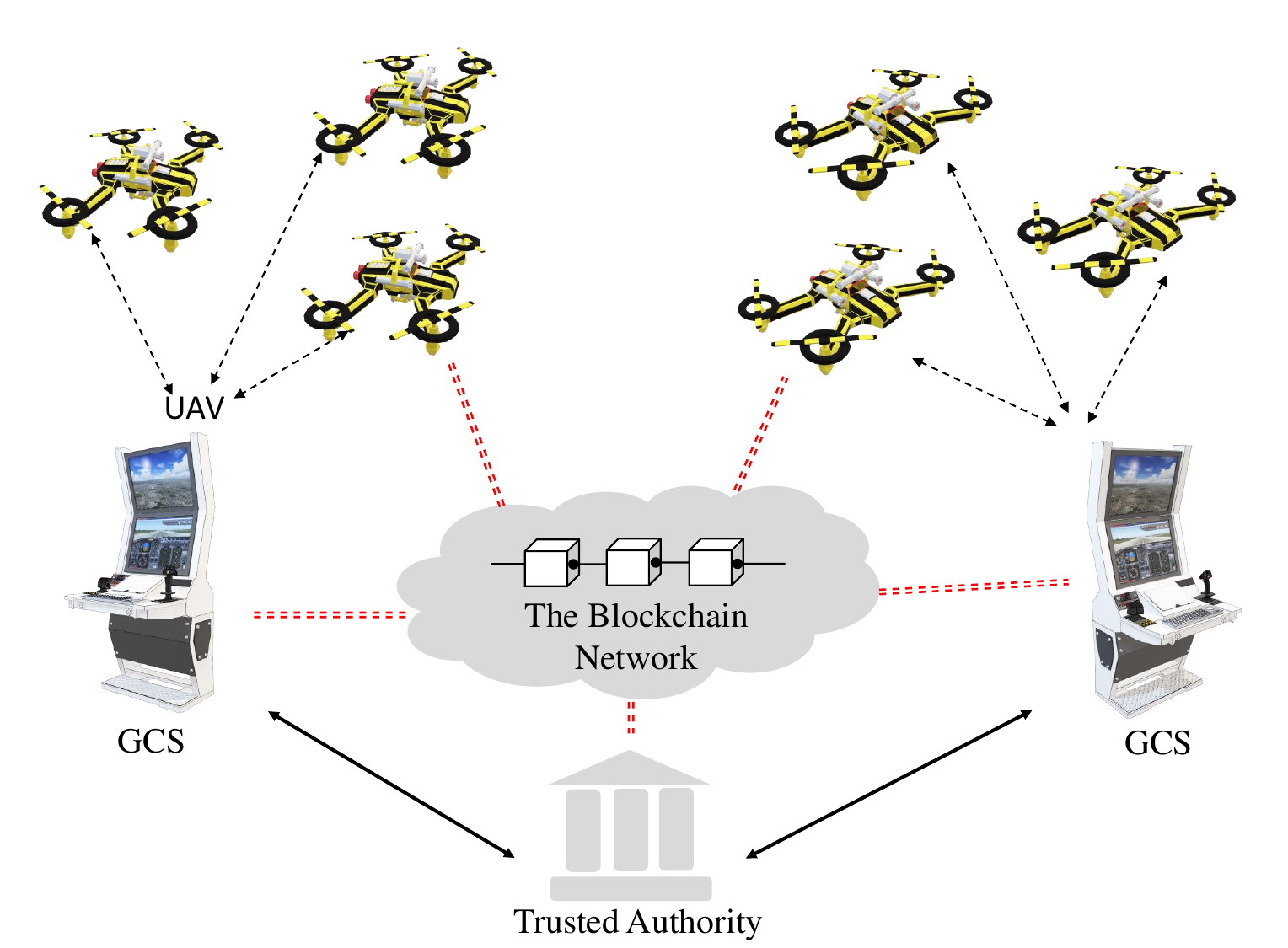}}
\setlength\abovecaptionskip{0.5\baselineskip}
\setlength\belowcaptionskip{-0.5cm}
\caption{UAV Ad-hoc Network}
\label{adhoc}
\end{figure}
\subsection{Scheme modeling}
\subsubsection{ Trusted authority (TA) }
The Trusted Authority is a trusted third party for key distribution. The $TA$ provides the secret keys $sk$ for Identity-based encryption schemes. The approved node responsible for monitoring other nodes' behavior or cooperation pattern is known as a $TA$ node which validates the identification of a UAV that intends to send messages or produces a new identity and verifies that another UAV possesses the specified identity.
\subsubsection{Smart Contract Deployer} 
The evidence also exists that smart contracts must establish a user account on the consortium blockchain. Therefore, to eliminate the trust barriers between domains, a $t, n$ threshold multi-signature smart contract is created. Let $n$ be the number of participants and $t$ the threshold. The number of elements contained in the above merkle tree is the combinatorial number $C(n, t)$. The space complexity of this tree, $O\left(n^{\wedge}(n-t)\right)$, is exponential on the threshold, and the complexity is $O(\log (C(n, t)))<O\left(\log \left(n^{\wedge} t\right)\right)=O(t \cdot \log (n))$.
\subsubsection{Ground Control Station (GCS)}
GCS receives UAV data, processes it, and converts and transfers it to other communication protocols to link clients on the same network for decisive piloting and communication between a UAV and its network. In addition, this GCS usually allows for UAV autopilots and live video and data streaming. Ground stations for UAVs are essential to a new era of long-range aerial data collection.
In recent years, reliable and secure communication has been scarce, The link between GCS and the UAV has been experienced, which is also a significant concern in our work.
\subsection{ BETA-UAV: The proposed Blockchain-based Efficient Authentication scheme}
BETA-UAV comprises three phases that can be described as follows.
\begin{enumerate}
    \item \emph{System initialization phase}: \\TA initializes the the system parameters as follows.
     TA initializes the elliptic curve $E: y^{2}=x^{3}+a x+b \bmod p$ such that $a, b \in Z_{q}^{*}$, $\Delta=4 a^{3}+27 b^{2} \neq 0$ \cite{9523518}, and $p, q$ are 160-bits prime numbers with 80-bits security. Based on the generator $g$, TA creates the cyclic group $\mathbb{G}$ that includes the points of $E$ in addition to the point of the infinity $\mathcal{O}$. TA selects the system secret parameter $Sk_{TA} \in Z_{q}^{*}$, then calculates its associated public parameter $Pk_{TA}=Sk_{TA}.g$. Secure hash function $H_{1}$, e.g., SHA-256. TA deploys the smart contract $SC$ through transaction $Tx$ and retrieves the $SC$'s address $SCID$. At last, the public parameters of the scheme are $PPs=\left\langle a, b, p, q, g, S C I D, H_{1}, Pk_{TA}\right\rangle$.
    \item \emph{Registration phase}:\\ For all the terminals in the network, TA is responsible for registering all GCSs and UAVs before being part of the network as follows.
    For each GCS $GCS_{j}$, TA creates a long-term digital certificate $Cert_{GCS_{j}}$ by selecting $Sk_{GCS_{j}} \in Z_{q}^{*}$, calculating $Pk_{GCS_{j}}=Sk_{GCS_{j}}.g$, and signing it to generate $\sigma_{TA}=Sign_{Sk_{TA}}(Pk_{GCS_{j}} \| T_{R})$, where $T_{R}$ is the expiration date. At last, $Cert_{GCS_{j}}=\left\langle Pk_{GCS_{j}}, T_{R}, \sigma_{TA} \right\rangle$.
    Similarly, for each unmanned aerial vehicle $UAV_{i}$, TA creates its long-term digital certificate as $Cert_{UAV_{i}}=\left\langle Pk_{UAV_{i}}, T_{R}, \sigma_{TA} \right\rangle$. 
    At last, TA loads $PPs$ and the certificate revocation list $CRL$ of revoked terminals onto all registered terminals as well as its issued digital certificate and secret key.
    \item \emph{Signature generation and verification phase}:\\
    Considering $UAV_{1}$ in the communication range of $UAV_{2}$, in this case, the authentication process is divided into authentication for the first and subsequent transmission slots as follows.\\
    For the first transmission slot:\\
        $UAV_{1}$ sends $UAV_{2}$ a communication request in the form of the tuple $\langle Cert_{UAV_{1}}, T_{1}, T_{S}, \sigma_{1} \rangle$, where $\sigma_{1}=Sign_{Sk_{UAV_{1}}}(Cert_{UAV_{1}} \| T_{1} \| T_{S})$ signed at $T_{1}$ timestamp and $T_{S}$ is the whole session time interval, e.g., [00:10:00].
        $UAV_{2}$ in turn checks $T_{1}$'s freshness, verifies $\sigma_{1}$ as $Verf_{Pk_{UAV_{1}}}(\sigma_{1})$, then triggers the Issue-$UAV_{2}$($Pk_{UAV_{1}}$, $Pk_{UAV_{2}}$, $T_{1}$) function in the smart contract using $SCID$ and retrieve $TxID_{2}$. At last, $UAV_{1}$ stores $\langle Pk_{UAV_{1}}, T_{S}, TxID_{2} \rangle$.
        Similarly, $UAV_{2}$ sends $UAV_{1}$ a reply in the form of the tuple $\langle Cert_{UAV_{2}}, T_{2}, T_{S}, \sigma_{2} \rangle$, where $\sigma_{2}=Sign_{Sk_{UAV_{2}}}(Cert_{UAV_{2}} \| T_{2} \| T_{S})$ signed at $T_{2}$ timestamp.
        $UAV_{1}$ in turn checks $T_{2}$'s freshness, verifies $\sigma_{2}$ as $Verf_{Pk_{UAV_{2}}}(\sigma_{2})$, then triggers the Issue-$UAV_{1}$($Pk_{UAV_{2}}$, $Pk_{UAV_{1}}$, $T_{2}$) function in the smart contract using $SCID$ and retrieve $TxID_{1}$. At last, $UAV_{2}$ stores $\langle Pk_{UAV_{2}}, T_{S}, TxID_{1} \rangle$.
        For subsequent transmission slots:
        For each message $m$, $UAV_{1}$ signs $m$ at $T_{3}$ timestamp to get $\sigma_{3}=Sign_{SK_{UAV_{1}}}(m \| T_{3} \| Pk_{UAV_{1}})$ and sends $\langle m, T_{3}, Pk_{UAV_{1}}, \sigma_{3} \rangle$ to $UAV_{2}$.
       $UAV_{2}$ checks $T_{3}$'s freshness, verifies $\sigma_{3}$ as $Verf_{Pk_{UAV_{1}}}(\sigma_{3})$, retrieves the $TxID_{2}$ related to the received $Pk_{UAV_{1}}$, and acquires $TxID_{2}$'s information from the blockchain to check the session continuity by finding out if $T_{3}-T{1} \leq T_{S}$ holds or not. If holds, $m$ will be accepted. Otherwise, it will be discarded.
\end{enumerate}
\section{Security Analysis}\label{SA}
The GCS and drone had certificates for registration from the TA. Both parties exchange credentials and check the authenticity of the certificates as $Cert_{UAV_{i}}=\left\langle Pk_{UAV_{i}}, T_{R}, \sigma_{TA} \right\rangle$ during the significant agreement process. Consequently, if the drone and ground station have valid certificates, they can authenticate each other.
\subsection{Message authentication}
$UAV_{1}$ sends $UAV_{2}$ a communication request in the form of a tuple $\langle Cert_{UAV_{1}}, T_{1}, T_{S}, \sigma_{1} \rangle$, where $\sigma_{1}=Sign_{Sk_{UAV_{1}}}(Cert_{UAV_{1}} \| T_{1} \| T_{S})$ is signed at $T_{1}$ timestamp and $T_{S}$. The intended recipient and receiver UAV share a symmetric key $S_{K}$ to determine the authentication process.
\subsection{Security protection against active attacks}
An attacker A can quickly monitor and eavesdrop on communication messages on a public channel if every message refreshes every session like $\sigma_{2}$ as $Verf_{Pk_{UAV_{2}}}(\sigma_{2})$, rendering it impractical for an attacker to extract all pertinent information.The BETA sends no parameters twice, so our protocol model resists tracking and eavesdropping.
\begin{enumerate}
\item \emph{Resilience to modification}:\\
Resilience is a fundamental requirement for multi-UAV operation. Because these systems operate in a dynamic and open environment, they are susceptible to various interruptions. For each message $m$, $UAV_{1}$ signs $m$ at $T_{3}$ timestamp to get $\sigma_{3}=Sign_{SK_{UAV_{1}}}(m \| T_{3} \| Pk_{UAV_{1}})$ and sends $\langle m, T_{3}, Pk_{UAV_{1}}, \sigma_{3} \rangle$ to $UAV_{2}$. A multi-UAV system is robust if it can accomplish the original mission at an acceptable level of performance, despite diversion.  
\item \emph{Resilience to replaying}:\\ The UAV assigns public key $pk$ and secret keys $sk$ at each authentication. Information from the blockchain checks session continuity by determining whether $T_{3}-T{1} \leq T_{S}$ holds. If this fits, $m$ is accepted. Otherwise, it was discarded directly.  
\item \emph{Resilience to impersonation}: \\ When an adversary $\mathbb{A}$ attempts to impersonate an unauthorized drone (e.g., Alice) he is required to compute a valid signature for a coherent topic using Alice's credentials.
Nonetheless, it is difficult for a Ts opponent owing to the message authentication characteristic, namely the  $T_{2}$'s freshness, to authenticate $\sigma_{2}$ as $Verf_{Pk_{UAV_{2}}}(\sigma_{2})$, and then trigger the ($Pk_{UAV_{2}}$, $Pk_{UAV_{1}}$, $T_{2}$) function in the smart contract using $SCID$ and retrieve $TxID_{1}$. Finally, $UAV_{2}$ stores $\langle Pk_{UAV_{2}}, T_{S}, TxID_{1} \rangle$.
\item \emph{Man-in-the-middle (MITM) Attack}:\\ Per a schema, an adversary can capture and compromise all messages sent and received $w$. The message exposure during the freshness identification process is  $\langle Cert_{UAV_{1}}, T_{1}, T_{S}, \sigma_{1} \rangle$,$\langle Cert_{UAV_{1}}, T_{2}, T_{S}, \sigma_{2} \rangle$.
If $\mathbb{A}$ attempts to reconstruct UAV certification, the contents of $UAV_{1}$ and $UAV_{2}$ must be modified. Moreover, for $\mathbb{A}$ to reconstruct UAVs, $pk$ and $Sk$ must be known; $pk$ and $Sk$ are the required parameters for message regeneration.

Hence, without requisite secret credentials, it is impractical for $\mathbb{A}$ to reissue a valid message. Therefore, BETA-UAV is resistant to MITM attacks.
\item \emph{Resilient to birthday collision:}\\ Our proposal could encounter this property if the endorsed blockchain is susceptible to birthday collisions. For our design, we employed developed blockchain systems, such as Ethereum, that support smart contracts. This distributed ledger system uses secure hash functions such as SHA-256 \cite{b9}.
Therefore, computing the block hash can eliminate the generation of two-birthday collision blocks.
\end{enumerate}
\section{Implementations and Performance Analysis}\label{IPA}
Our BETA-UAV protocol demonstrates its prototype blockchain implementation in Ethereum test networks, its demonstrated efficiency in drone authentication, and a simulated UAV ad hoc network scenario. Performance is then considered in the context of the implementation outcomes. 
\subsection{Implementations}
First, we deployed our smart design contract on an online public Ethereum test network (Rinkeby Test Network). Rinkeby offers a comprehensive development environment ID for proficiently compiling and deploying solid smart contracts. This expedites the prototyping process for blockchain-enabled systems. Specifically, we employed the following Remix settings compiler (0.8.7. commit.228d28d7).
Our gas cost analysis begins by compiling our Solidity Smart Contract Code, which is subsequently deployed in the configuration described above using Remix. The first is the gas price of Eth, which reflects the cost of maintaining an Ethereum blockchain \cite{b7}. we simulate cryptographic primitives in desktop and Raspberry PI environments with the configurations" Linux Ubuntu 18.04 LTS, Intel Core Processor 11th Gen Intel(R) Core(TM) i7-11850H @ 2.50GHz; we simulate cryptographic primitives in desktop and Raspberry Pi environments with the configurations" Linux Ubuntu 18.04 LTS, Intel Core CPU @ 3.60 GHz, 8 GB RAM" and" Raspberry PI 4B, Quad-core ARM Cortex-A72 @ 1.5 GHz, 16GB RAM", respectively\cite{9760471}.
\begin{table}[!t]
\centering
\caption{ Remix settings.}
\begin{tabular}{|c|c|}
\hline
\makebox[1em]{Parameter}     & \makebox[-0.5em]{Value}             \\ \hline\hline
Compiler               & 0.8.7.commit.228d28d7      \\ \hline
Language               & Solidity                   \\ \hline
EVM version            & Compiler default          \\ \hline
Deployment Environment & JavaScript Virtual Machine \\ \hline
Featured Plugins &
 \begin{tabular}[c]{@{}c@{}}Solidity Compiler, \\Deploy and Run Transactions\\ Solidity Static Analysis,\\ and Solidity Unit Testing
 \end{tabular} \\ \hline
\end{tabular}
\label{settings}
\end{table}
\subsection{Computation Cost Comparison}
Compared to prior schemes by the authors [8]–[11], the BETA-UAV performance IoD was determined based on their computational and communication costs. For the experimental examination of various cryptographic primitives, we implement the widespread "Multi-precision Integer and Rational Arithmetic Cryptographic Library (MIRACL)."  
Therefore, using the MIRACL library, we simulated and evaluated the execution times of cryptographic primitives [7]. This section determines the computational cost for the proposed scheme and the associated schemes. The simulation results are listed in Table 1, and the total computational costs of our scheme and other related schemes are listed in Table 2.

According to the results, the proposed scheme has higher computational efficiency than the other schemes.
As shown in Table III, BETA-UAV requires a lower computational cost on the user side than related existing schemes [8]–[11]. 
GCS, stationed at the TA, is an essential component of the UAV environment. Consequently, it is preferable to reduce the computational cost of the central server. The computational cost at the CS side in the proposed BETA-UAV is  [19.28] ms, whereas [8]-[11] require 
$5 \tau_{H F}+3 \tau_{E P M}+\tau_{F E} \approx[0.848] \mathrm{ms}$. $9 \tau_{H F}+3 \tau_{E P M}+2 \tau_{E P A} \approx[2.084] \mathrm{ms}$ $1 \tau_{H F}+5 \tau_{E P M}+\tau_{E P A} \approx[3.058] \mathrm{ms}$ and $11 \tau_{H F}+$ $3 \tau_{E P M}+1 \tau_{E P A} \approx[2.138] \mathrm{ms}.$ 

Therefore, BETA-UAV has a lower computational cost than the schemes shown in Table 2. Even so, BETA-UAV has a lower computational cost than the alternative schemes. In contrast to the other schemes, the BETA-UAV has a lower computational cost $17 \tau _{HF}+7 \tau_{I M A}+ $5 +$\tau_{H F} \approx[19.28] \mathrm{ms}.$ is the computational cost of the drone (Dx) or sensor node in the proposed BETA-UAV, whereas [8]–[11] requires $8 \tau_{H F}+$ $4 \tau_{E P M}+1 \tau_{E P A} \approx[14.38] \mathrm{ms}.$$12 \tau_{H F}+4 \tau_{E P M}+\tau \approx[16.32] \mathrm{ms}.$
For the transaction hash: $\quad 0x28ef49323cafc471a9a7d5\ldots$.
\begin{table}[!t]
  \centering
    \begin{tabular}{|l|c|c|l|}
\hline   Notations & Primitives & PF-1 & PF-2 \\
\hline  $\tau_{I M A}$ & Instance multiplication & $2.79 \mathrm{~ms}$ & $0.602 \mathrm{~ms}$ \\
\hline  $\tau_{I P A}$ & Instance point addition & $0.003 \mathrm{~ms}$ & $0.145 \mathrm{~ms}$ \\
\hline  $\tau_{H F}$ & Hash Functions & $0.301 \mathrm{~ms}$ & $0.029 \mathrm{~ms}$ \\
\hline  Ts & Timestamp & $1.0 .1 \mathrm{~ms}$ & $1.1 .1 \mathrm{~ms}$  \\
\hline $\tau_{E N C}$ & SHA-256 & $0.485 \mathrm{~ms}$ & $0.085 \mathrm{~ms}$  \\
\hline
\end{tabular}
\end{table}
\subsection{Estimate Gas Cost}
Ethereum undergoes simple computations that coincide with a swarm of computers called nodes. An elite group of nodes is defined as the miners who work the hardest. Miners protect the network from intrusion and prioritize the computations. Therefore, the miners must pace a stream of requests. Without this, the network might become overloaded owing to heavy usage or spammers picking up what is done. First, miners rely on the gas price, and the gas limit of the last unit measures the work, but it has no monetary value; miners pay in tiny denominations of ETH called Gwei. In this study, we deployed a smart contract to a rinkeby test network. We then connected and deployed it to the meta mask. Once the transaction is confirmed and mined, we go to the blockchain explorer page to see the number of gas units used for this transaction. 
For the transaction hash: $\quad 0x28ef49323cafc471a9a7d5\ldots$
The Gas Price is shown in Fig. 2 as follows: $0.000000002566484836 \text{ Ether } (2.566484836 \text{ Gwei})$.

\begin{table}[htb!]
\caption{Comparison of actual vs estimated cost.}
\centering
\begin{tabular}{|c|c|c|}
\hline
\textbf{Function} & \textbf{Estimated} & \textbf{Actual} \\
\hline
Deployer & 0.0005499 ETH & 0.000555 ETH \\
\hline
Issue UAV1 & 0.00023767 ETH & 0.000238 ETH \\
\hline
\end{tabular}
\label{tab:cost_comparison}
\end{table}

\begin{figure}[t!]
\centerline{\includegraphics[width=9cm]{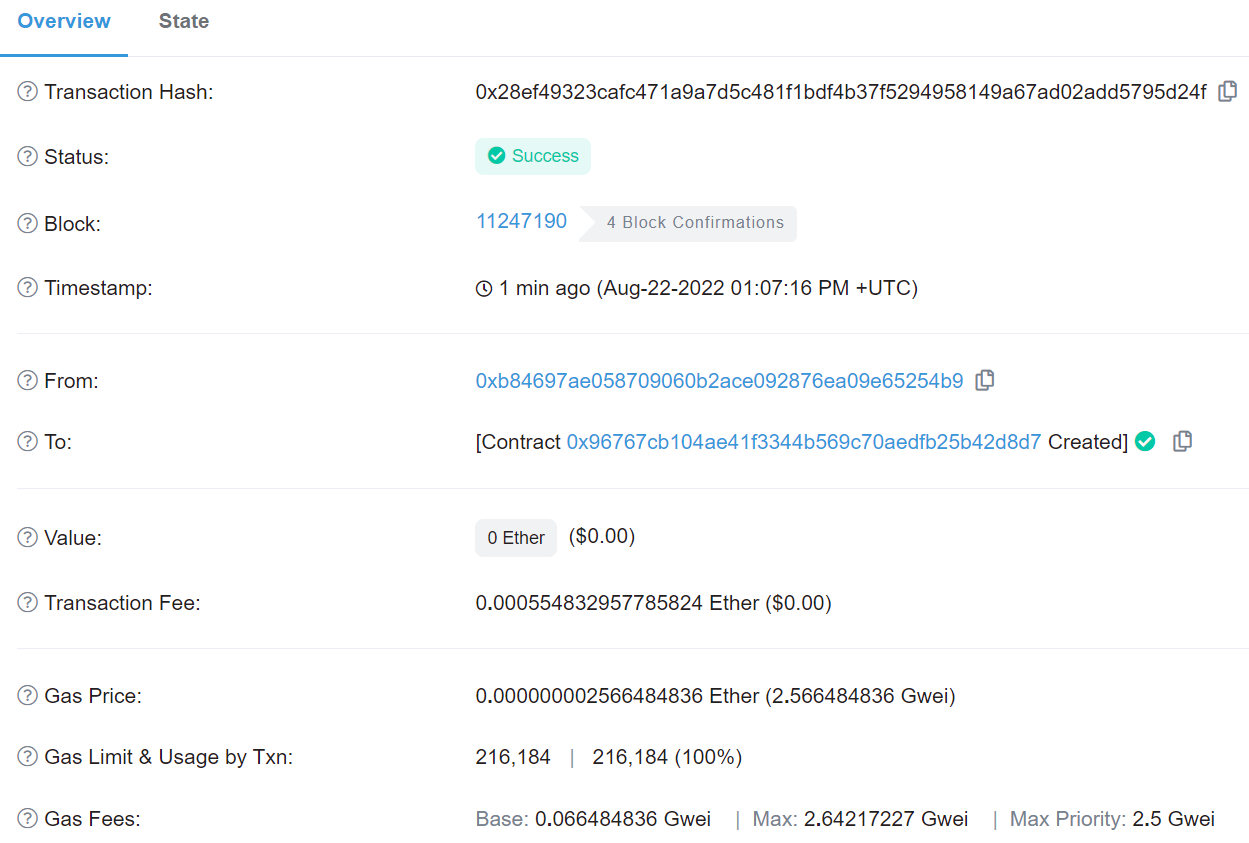}}
\setlength\abovecaptionskip{0.5\baselineskip}
\setlength\belowcaptionskip{-0.5cm}
\caption{Network Deployment}
\label{ND}
\end{figure}

\begin{figure}[t!]
\centerline{\includegraphics[width=8.5cm]{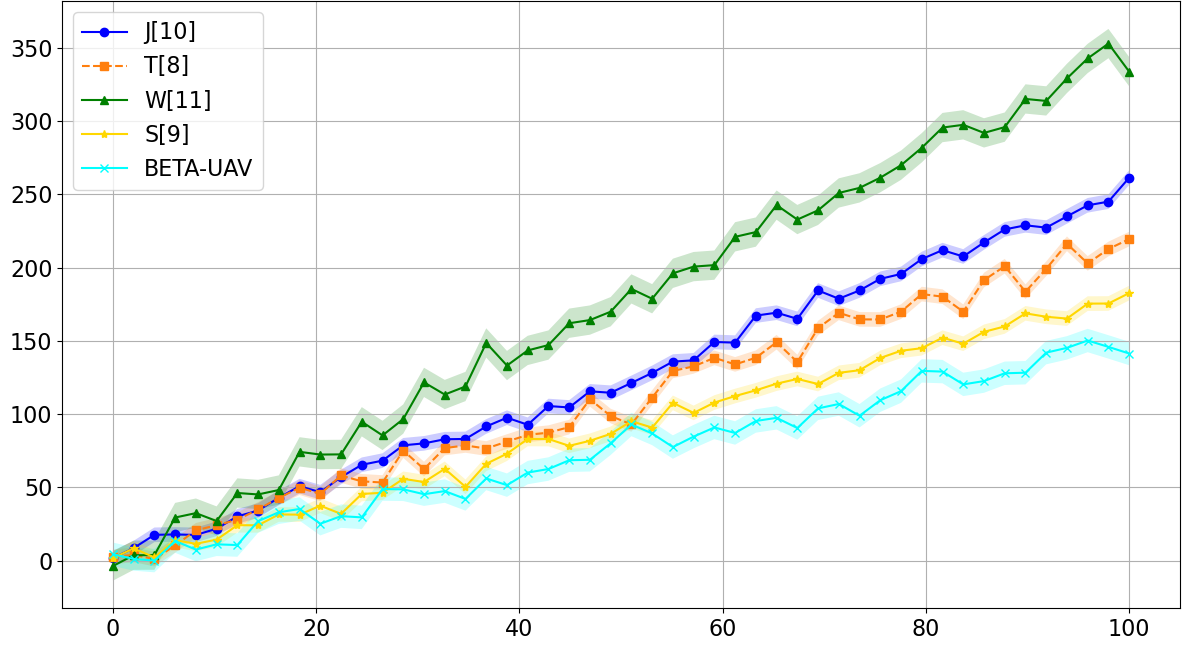}}
\setlength\abovecaptionskip{0.5\baselineskip}
\setlength\belowcaptionskip{-0.5cm}
\caption{Computational Delay vs Number of Drones.}
\label{f1}
\end{figure}

\begin{figure}[t!]
\includegraphics[width=9cm]{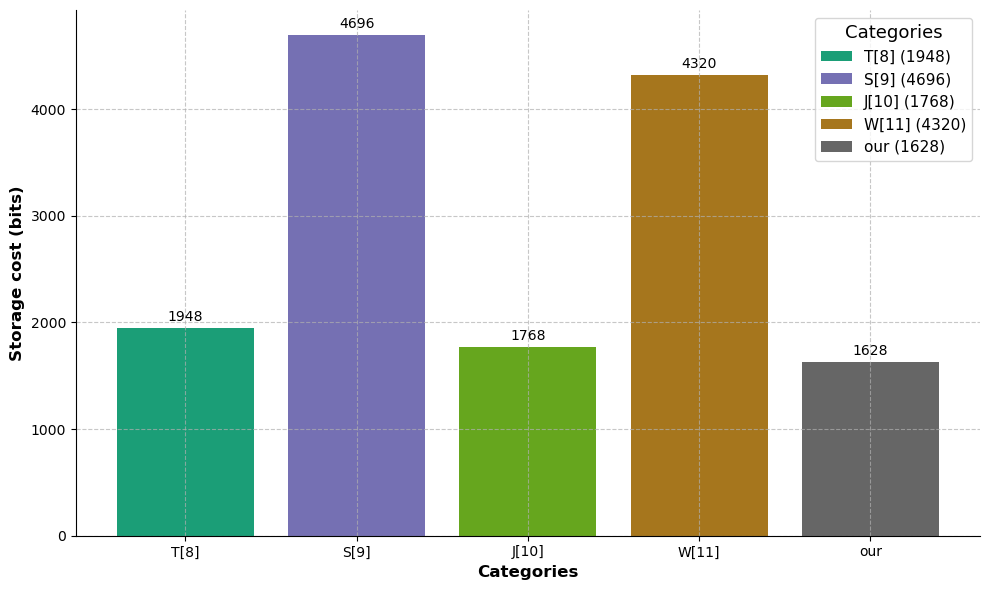}
\caption{Comparison of Computational Cost }
\label{fig:CC}
\end{figure}
In Fig .\ref{f1} BETA-UAV system demonstrates the most feasible and efficient computational delay performance in this graph, with low baseline delay that scales gradually and predictably with number of nodes. This makes it the most promising system overall based on the results visualized.
\subsection{Communication Cost Comparison}
We evaluated our scheme's communication costs compared with the existing algorithms discussed above. The identity, hash function, random number, SHA-256, timestamp, and modular exponentiation are respectively $32$ bits, $256$ bits, $160$ bits, and $128$ bits. We procure the communication cost of the proposed scheme for each message as follows: $2240$ bits, $3360$ bits, $2656$ bits, and $3200$ bits by applying these notations.
Therefore, the proposed scheme has a total communication cost of $160 + 256 + 40 + 100 \approx 556 \, \mathrm{bits}$.

In this section, we compare the communication costs of the proposed protocol with those of the related schemes [8]–[11]. The outcome indicates that the proposed method has lower communication costs than existing solutions.
The bar chart Fig. \ref{fig:CC} compares the storage costs in bits for different works. The costs range from 1628 bits for our own data to 4696 bits for category S[9]. The legend shows the exact storage cost for each category. Overall, the graph illustrates the relative storage requirements for the data categories, with S[9] being the most expensive and our data being the most efficient. This comparison highlights the storage optimization achieved for our method.
\section{Conclusion}\label{Con}
In this study, we proposed a blockchain-based efficient authentication scheme called BETA for UAV communication, where BETA-UAV is divided into three phases: registration, authentication, and signature verification.
The routing framework can endure major security attacks based on informal security analysis. Our study aims to address this security vulnerability by proposing a provable efficient authentication scheme that protects user privacy. Significant advantages are identified in the proposed scheme, such as lower computational and communication costs, small key size, and greater secrecy.
In the future, we will extend this technical work to Ethereum cryptography and compare it with the encryption algorithms. Furthermore, we will implement the computational costs of the proposed work in a practical scenario. 
\balance

\vspace{12pt}
\end{document}